\documentstyle[aps,prb]{revtex}
\begin{document}
\draft
\twocolumn[\hsize\textwidth\columnwidth\hsize\csname @twocolumnfalse\endcsname

\title{Reply to Comment  "Invalidity of classes of approximate Hall
effect calculations." }
\author{N.B. Kopnin and  G.E. Volovik}

\address{Low Temperature Laboratory, Helsinki University of
Technology, 02150 Espoo, Finland, \\ and \\ Landau Instute for
Theoretical
Physics,
117334
Moscow, Russia}

\date{\today}
\maketitle

\

]
\narrowtext

In his Comment \cite{Ao} to our paper \cite{KopVol} Ao points out the main
reason why he thinks
our calculations are not correct. The reason is not specific to the
particular paper \cite{KopVol} but rather refers to all our
microscopic calculations (see also \cite{KopKrKL}, etc.) of vortex dynamics.
It is, as he claims, the incorrect use of the $\tau$-approximation.
Ao agrees that the $\tau$-approximation works well for calculations of
conductivity but states, strangely enough, that it fails when
applied for calculating resistivity. Instead, Ao suggests to use the approach
developed in Ref. \cite{Ao1}. Being intrigued by the possibility to discover
the truth, we turn to Ref. \cite{Ao1} and find that the starting point is
exactly the same as in all our work on the subject: one looks for a response
of the superconductor to a time-dependent displacement of the vortex, i.e.,
the problem under consideration is exactly the calculation of the
conductivity tensor $\sigma _{xx}$ and $\sigma _{xy}$ in the mixed state.
Why then one needs
to speak of a misteriuos unsufficiency of the $\tau$-approximation?
The further analysis of Ref. \cite{Ao1} shows that all  the calculations refer
to the case when the relaxation time is
infinite. One would then expect that the results of Ref. \cite{Ao1} coincide
with ours without impurities. And they really do in the limit
$\tau \to \infty$ but only for zero
temperature. For a finite temperature, however, there is no Iordanskii
force in Ref. \cite{Ao1}. The reason is easily seen if one follows
how the Bogoliubov-de Gennes equation in Ref. \cite{Ao1} is transformed
into Eq. (16). Here the  phase factor $\exp 2i(\delta _k-\delta
_{k^\prime})$ is
lost, where
$\delta _k$  is the scattering phase-shift for a particle travelling near a
vortex.  It is this
phase factor which determines both the transverse and longitudinal
scattering cross sections \cite{KopKr1}, the former being responsible for the
Iordanskii force \cite{Sonin}. Turning to a finite relaxation,
Ao speculates that it should affect (!) the longitudinal vortex response  but
would not change (?) the transverse response. The argumentation is based on
Eq. (15) which has been derived in absence of impurities! The transformation
uses the Schr\"odinger equation which does not include interaction with either
impurities or other sources of relaxation.  Thus there is the important step
missing in Ao's calculations: one has to incorporate the interaction with
impurities into the starting equations and proceed carefully to get the correct
result:  The conductivity (not resistivity!)
roughly obeys the semicircle law in an isotropic superconductor
\cite{KopKrKL}:
$\sigma_{xx}\propto
\omega_0\tau/(1+\omega_0^2\tau^2)$~,~$\sigma_{xy}\propto
\omega_0^2\tau^2/(1+\omega_0^2\tau^2)$. The crossover is determined by the
lowest energy scale in superconductor: the minigap $\omega_0$ of core fermions.
That is why the
Hall conductivity $\sigma_{xy}$ is extremely small in the dirty limit and
in the Ginzburg-Landau theory, where $\omega_0 \tau \rightarrow 0$, -- the
points which Ao theory fails to explain.

Ao presents handwaving arguments that everything is wrong
which is not as simple as he wants:
(1) Experiments in which the forces on $^3$He vortices have been measured
in a wide temperature range \cite{Bevan} are too complicated and thus can be
wrong;
(2) Microscopic calculations using the Green function formalism \cite{KopKrKL}
is wrong;
(3) The spectral flow pheneomenology in terms of the Landau-type theory
for the Fermi system in vortex cores \cite{Stone} uses $\tau$-approximation
and thus is wrong, etc. Then there is a puzzle: why all three
sources agree in the
temperature dependence of both longitudinal and Hall conductivities?

The confusion regarding the transverse  force on a vortex is typical
for those who start to consider this problem. Here the simple results
advocated by Ao can be dangerous: it is a  strong temptation to
take a simple formula and use it without precaution. But once the
importance of the quasiparticle transport in the core and outside
the core is realized, one can move further. One
finds a lot of interesting things on this way:  Landau damping
coming from the gapless excitations in $d$-wave superconductors
\cite{KopVol}; rotational dynamics of the nonaxisymmetric
vortices \cite{KopninVolovik1998};  relation to event
horizon\cite{KopninVolovik1998};  mesoscopic effects and Zener tunneling in
the core\cite{FeigelmanSkvortsov,Larkin-Ovchinnikov}; nonlinear
transport, etc.

There is a long way for Ao to go...

\end{document}